\documentclass[graybox]{svmult}

\usepackage{type1cm}        
\usepackage{makeidx}         
\usepackage{graphicx}        
\usepackage{multicol}        
\usepackage[bottom]{footmisc}

\usepackage{newtxtext}       %
\usepackage{newtxmath}       

\usepackage[normalem]{ulem}
\usepackage{xcolor}

\makeindex             

\begin{document}

\title*{Neutron Stars} 
\author{Jutta Kunz} 
\institute{Jutta Kunz \at Institute of Physics, University of Oldenburg, Germany, \email{jutta.kunz@uni-oldenburg.de}
}
%
%
\maketitle

\abstract*{Neutron stars are highly compact astrophysical objects and therefore of utmost relevance to learn about theories of gravity.
Whereas the proper equation of state of the nuclear matter inside neutron stars is not yet known, and a wide range of equations of state is still compatible with observations, this uncertainty can be overcome to a large extent, when dimensionless neutron star properties are considered.
In this case \textit{universal relations} between neutron star properties and for the gravitational radiation in the form of quasi-normal modes arise. 
These \textit{universal relations} can be rather distinct for alternative theories of gravity as compared to General Relativity.
Moreover, the presence of new degrees of freedom in alternative theories of gravity leads to further types of gravitational radiation, that may be revealed by pulsar observations and gravitational wave detectors. 
Here an introduction to neutron stars, their properties and \textit{universal relations} is presented, followed by two examples of alternative theories of gravity featuring interesting effects for neutrons stars.
}

\abstract{Neutron stars are highly compact astrophysical objects and therefore of utmost relevance to learn about theories of gravity.
Whereas the proper equation of state of the nuclear matter inside neutron stars is not yet known, and a wide range of equations of state is still compatible with observations, this uncertainty can be overcome to a large extent, when dimensionless neutron star properties are considered.
In this case \textit{universal relations} between neutron star properties and for the gravitational radiation in the form of quasi-normal modes arise. 
These \textit{universal relations} can be rather distinct for alternative theories of gravity as compared to General Relativity.
Moreover, the presence of new degrees of freedom in alternative theories of gravity leads to further types of gravitational radiation, that may be revealed by pulsar observations and gravitational wave detectors. 
Here an introduction to neutron stars, their properties and \textit{universal relations} is presented, followed by two examples of alternative theories of gravity featuring interesting effects for neutrons stars.
}

\label{sec:1}

\section{Introduction}

When massive stars with initial mass $M \gtrsim M_\odot$ have burnt the nuclear fuel in their core gravitational collapse results, leaving behind a highly compact remnant, a neutron star or a black hole. 
While predicted shortly after the discovery of the neutron \cite{Baade:1934zex}, neutron stars were only observed in the late 60s, when very regular radio pulses appeared in the data taken by Jocelyn Bell \cite{Hewish:1968bj}.
The radio pulses were emitted by a pulsar, now known as PSR B1919+21, a rapidly rotating neutron star with misaligned magnetic field. 
Ever since numerous pulsars including a double pulsar have been discovered \cite{Taylor:1993ba,Kramer:2003py,Burgay:2003jj,Fermi-LAT:2013svs,pulsars}.
The extreme regularity of these pulses allows for high precision tests of General Relativity and severe constraints for various alternative theories of gravity (see e.g. \cite{Stairs:2003eg,Kramer:2021jcw}).

On the theoretical side, Tolman, Oppenheimer and Volkoff (TOV) considered in the 30s already the description of neutron stars, deriving and solving the TOV equations for a simple equation of state (EOS) of the nuclear matter, namely a cold Fermi gas \cite{Tolman:1939jz,Oppenheimer:1939ne}.
This work had profound implications, since it showed that neutron stars can be supported against the gravitational pull only up to a maximum mass, while beyond this mass the collapse of the stellar core will continue and lead to a black hole.
The value of the maximum mass depends of course on the EOS for the nuclear matter.
The proper EOS for nuclear matter under such extreme conditions as present in neutron stars is still unknown, though \cite{Lattimer:2012nd,Ozel:2016oaf,Baym:2017whm}).

In recent years much progress has been made based on the discovery of gravitational waves and the advent of multi-messenger astronomy \cite{LIGOScientific:2016aoc,LIGOScientific:2017vwq,LIGOScientific:2017ync}.
In particular, the observation of GW170817, where the merger of a neutron star binary was reported and analyzed led to new constraints on the neutron star EOS, since it allowed to put constraints on the tidal effects experienced by the coalescing bodies and on the neutron star radii \cite{LIGOScientific:2018cki}.
Further analysis also hinted at a new value (range) for the maximum mass of neutron stars \cite{Alsing:2017bbc,Rezzolla:2017aly}.
Previous observations of pulsars had already revealed the existence of neutron stars with masses of about 2 solar masses \cite{Demorest:2010bx,Antoniadis:2013pzd,NANOGrav:2019jur}.

In the following we will focus mainly on static neutron stars.
We will start with the derivation of the TOV equations,
and then address a set of important neutron star properties.
Besides their mass and radius, we will consider their moment of inertia, their rotational quadrupole moment and their tidal deformability.
Subsequently we will address seismology of neutron stars.
Thus we will consider the normal modes and quasi-normal modes (QNMs) of neutron stars, representing their reaction to perturbations.
The uncertainty of the EOS reflected in the neutron star properties and QNMs will then be largely reduced with the help of universal relations (see e.g. \cite{Yagi:2016bkt,Doneva:2017jop}).
Our final concern will be the consideration of neutron stars in a set of alternative gravity theories featuring an additional scalar degree of freedom, where we will highlight some interesting new aspects as compared to General Relativity.

\section{Static Neutron Stars}

\subsection{Tolman-Oppenheimer-Volkoff Equations}

In General Relativity (GR) static neutron stars are obtained by solving the Tolman-Oppenheimer-Volkoff (TOV) equations for a given equation of state (EOS) of the nuclear matter.
We will now derive this set of equations.

To this end we start from the Einstein equations
\begin{equation}
G_{\mu\nu}= {\cal R}_{\mu\nu}-\frac{1}{2}g_{\mu\nu}{\cal R}=
{8 \pi G} T_{\mu\nu}  ,
\end{equation}
and employ the stress-energy tensor of an isotropic perfect fluid describing the nuclear matter
\begin{equation}
T_{\mu\nu} = \left(\rho + P\right) u_\mu u_\nu +g_{\mu\nu} P ,
\end{equation}
whose four-velocity in the rest system is
$u^\mu = \left( u^t,0,0,0 \right)$.
In mixed co- and contravariant components the stress-energy tensor then reads
$T^\mu_{\ \nu} = {\rm diag}(-\rho , \ P  , \ P  , \ P  \ ) $
with energy density $\rho$ and pressure $P$.

A convenient metric ansatz for static spherically symmetric neutron stars is given by
\begin{equation}
ds^2 = g_{\mu\nu} dx^\mu dx^\nu
= -e^{2\Phi(r)} dt^2 + 
e^{2\Lambda(r)} dr^2
+ r^2 \left( d\theta^2 + \sin^2\theta d\phi^2 \right) ,
\end{equation}
which contains two unknown functions, $\Phi(r)$ and $\Lambda(r)$.
Expressing the metric function $\Lambda(r)$ in terms of the mass function $m(r)$
\begin{equation}
e^{2\Lambda(r)} = \frac{1}{1 -\frac{2 m(r)}{r}} ,
\end{equation}
the Einstein Tensor becomes with this ansatz 
\begin{eqnarray}
G_{00} & = & e^{2\Phi}\frac{2 m'}{r^2}  , 
 \\ 
G_{rr}  & = & \frac{2}{r} \left( \Phi' -
                        \frac{m}{r^2} \left(1-\frac{2m}{r}\right)^{-1}
                           \right) ,
\\ 
G_{\theta\theta} & = & r^2 \left[
\left(\Phi'' + \Phi'^2\right) \left(1-\frac{2m}{r}\right)
+\frac{\Phi'}{r} \left(1-m' -\frac{m}{r}\right)
-\frac{1}{r^2}\left(m' -\frac{m}{r}\right) \right] ,
\\ 
G_{\varphi\varphi} & = & \sin^2\theta G_{\theta\theta} .
\end{eqnarray}

From the Einstein equations
 $G^0_0 = \kappa T^0_0 $ and 
$  G^r_r = \kappa T^r_r$ ($\kappa=8 \pi G$)
we find
\begin{eqnarray}
m'  & = & \frac{\kappa}{2} \rho r^2 , \label{*}
\\
\Phi' & = &\frac{\frac{\kappa}{2}r^3 P + m}{\left(1-\frac{2m}{r}\right)r^2}
. \label{**}
\end{eqnarray}
Employing these two equations in the
Einstein equation
$G^\theta_\theta = \kappa T^\theta_\theta$
using
\begin{equation}
\Phi'' = \frac{d}{dr}\Phi' = 
\frac{d}{dr}\left(
\frac{\frac{\kappa}{2}r^3 P + m}{\left(1-\frac{2m}{r}\right)r^2}
\right) 
\end{equation}
we obtain the equation for pressure $P$,
where we can identify the
Newtonian part (underlined)
and the relativistic corrections
\begin{equation}
\underline{P' = -\frac{m \rho }{r^2}\left(1 + \frac{P}{\rho}\right)}
\left(1 +\frac{\kappa}{2} \frac{P}{m} r^3  \right)
                        \left( 1-\frac{2m}{r} \right)^{-1} .
                        \label{***}
\end{equation}
The system of equations (\ref{*}), (\ref{**}) and (\ref{***}) are the \textbf{TOV equations}, representing three equations for four unknowns.
Therefore we have to provide an EOS $\rho(P)$ in order to solve the equations.
A relatively simple EOS is the so-called polytropic EOS
\begin{equation}
\rho = \frac{P}{\Gamma-1} +\left( \frac{P}{K} \right)^\Gamma ,
\label{poly}
\end{equation}
where $\Gamma$ is the adiabatic index and $K$ the polytropic constant.
Many realistic EOSs can be parametrized as piecewise polytropic EOSs (see e.g. \cite{Read:2008iy}).

Neutron stars are compact objects with a given radius $R$.
Outside this radius the pressure and the density vanish.
Therefore, the exterior is simply described by the Schwarzschild spacetime.
Asymptotic flatness requires that the function $\Phi$ satisfies $\Phi(\infty)=0$. 
The mass function $m(r)$ assumes its asymptotic value $M$ at the surface of the star, where $M$ corresponds to the mass of the neutron star in geometric units.
At  the center regularity requires that $m(0)=0$.
The density and the pressure at the center are $\rho_c(P_c)$ and $P_c=P(0)$, respectively.
$P_c$ is a free parameter. 

\begin{figure}[ht]
\begin{center}
\mbox{
\includegraphics[width=.65\textwidth, angle =270]{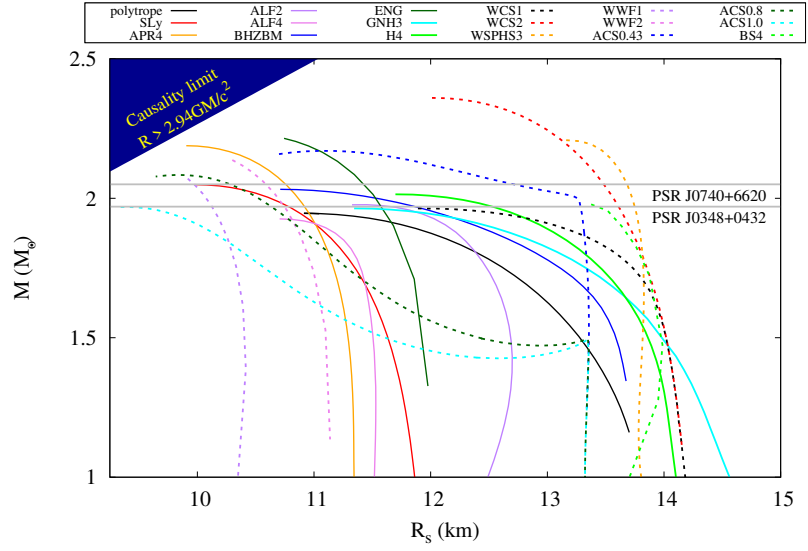}
}
\end{center}
\vspace{-0.5cm}
\caption{{\it{Mass-radius relation of neutron stars in GR: mass $M$ (in solar masses $M_\odot$) vs radius $R$ (in km) for numerous EOSs. 
The horizontal lines indicate high mass pulsars
\cite{Demorest:2010bx,Antoniadis:2013pzd,NANOGrav:2019jur},
the upper left corner marks the causality limit. From \cite{Blazquez-Salcedo:2022}.}}}
\label{fig1}
\end{figure}

By varying the central pressure
a family of neutron star solutions for a given EOS is obtained. 
The mass-radius relation of numerous such families of neutron stars is shown in Fig.~\ref{fig1}.
Clearly, there is a strong EOS dependence of the mass-radius relation. 
Observations of high mass pulsars constrain the EOSs, however, since their maximum mass should allow for the measured mass values \cite{Demorest:2010bx,Antoniadis:2013pzd,NANOGrav:2019jur}.

\subsection{Properties}

While the mass and radius of a neutron star are easily obtained, once an equation of state and a central pressure are specified, further properties of interest typically involve perturbation theory around the TOV background solution.
In lowest order perturbation theory the moment of inertia $I$ is obtained.
To this end, we consider a slowly rotating neutron star, that rotates with uniform angular velocity $\Omega$ around the axis $\theta=0$ ($\pi$). 
The metric then acquires a non-diagonal component
\begin{equation}
\delta g_{t\phi} = - \epsilon \omega\, r^2 \sin^2\theta ,
\end{equation}
where $\epsilon$ is a perturbation bookkeeping parameter, and the new metric function $\omega$ arising from the rotation needs to be determined.
All other rotational effects in the metric are of higher order.
This also holds for the effects on the density and pressure, which are even functions under time reversal.
The fluid velocity receives a contribution
\begin{equation}
\delta U^\mu = \left(0,0,0, \epsilon \Omega U^t \right) ,
\end{equation}
where $U^t= e^{-\Phi}$ is the time-component in the non-rotating frame.

\begin{figure}[ht]
\begin{center}
\mbox{\hspace{-0.3cm}
\includegraphics[width=.51\textwidth, angle =0]{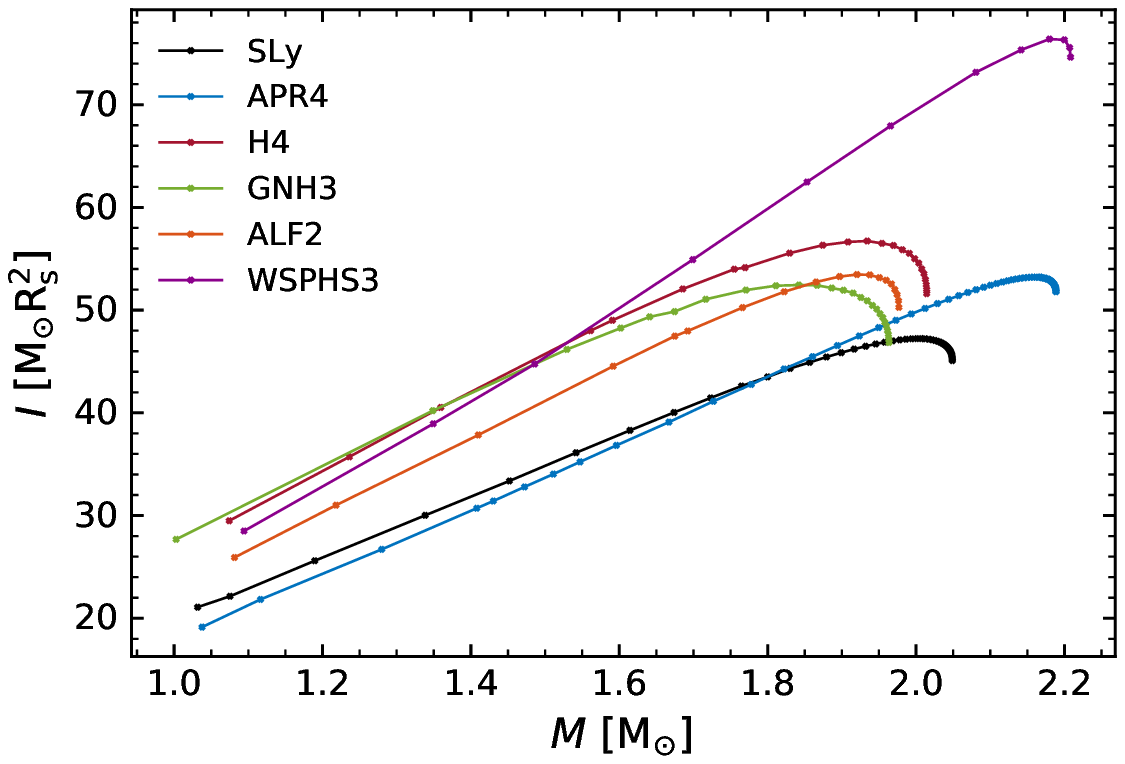}
\includegraphics[width=.51\textwidth, angle =0]{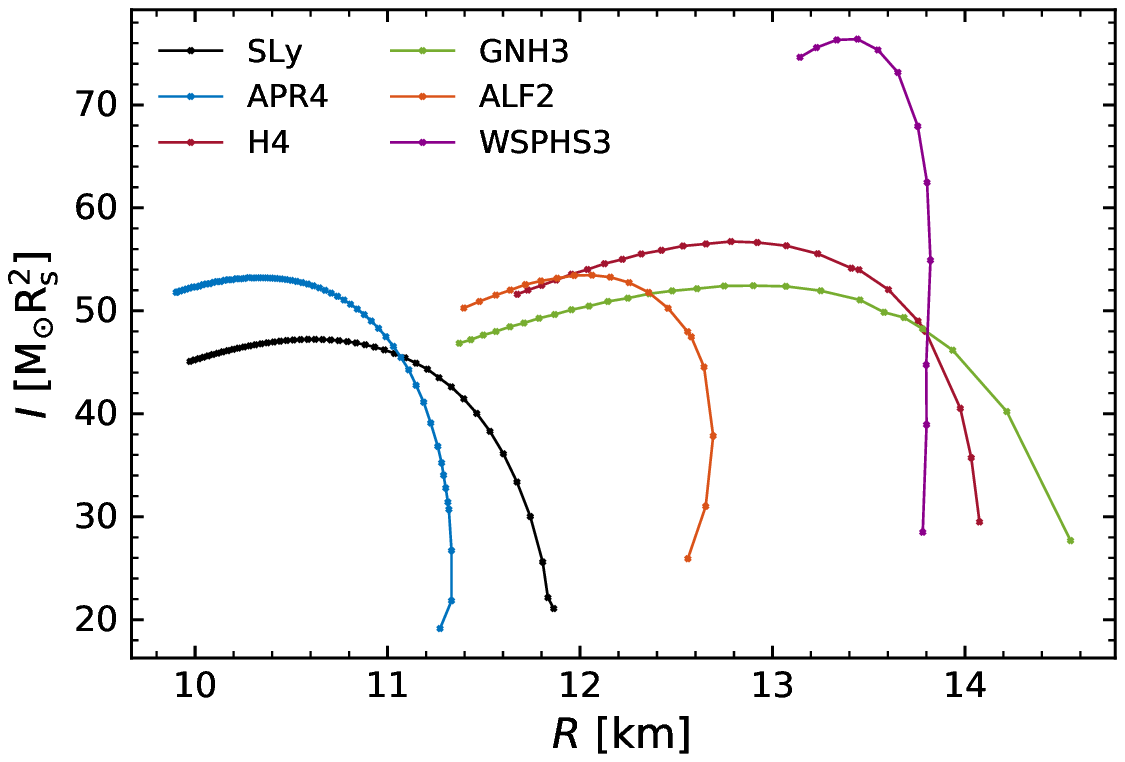}
}
\end{center}
\vspace{-0.5cm}
\caption{{\it{Moment of inertia $I$ of neutron stars in GR: (a) $I$ (in solar masses $M_\odot$ times the squared solar gravitational radius $R_S$) vs mass $M$ (in solar masses $M_\odot$); (b) $I$ vs radius $R$ (in km) for several EOSs. 
}}}
\label{fig2}
\end{figure}

The slow rotation induces a new component in the stress-energy tensor 
\begin{equation}
  \delta {T}_{t\phi} = r^2 \left(\rho+P\right)
       \epsilon \left(\omega-\Omega\right)\sin^2\theta 
       - P \epsilon\omega\, r^2 \sin^2\theta .
\end{equation}
A priori, the new function $\omega$ could depend on two coordinates, $r$ and $\theta$.
Moreover, the resulting partial differential equation does not separate, therefore an expansion in terms of vector spherical harmonics should be made \cite{Hartle:1967he,Sotani:2012eb}.
Inspection of the boundary conditions w.r.t.~regularity and asymptotic flatness shows, however, that only a single $l$ can contribute, $l=1$, leaving $\omega$ as a function of $r$ only, determined by
\begin{align}
  \omega'' + \left( \frac{4}{r} - \Phi' - \Lambda' \right) \omega' & - 2 \left[ \Phi'' + \left( \Phi' - \Lambda' \right) \left( \Phi' + \frac{1}{r} \right) \right] \omega  \nonumber \\
       & + 2 \kappa e^{2\Lambda} \left[ P\omega + \left( \rho
       + P \right) \left( \Omega-\omega \right) \right] = 0.  \label{Gt3phi}
\end{align}
Expansion at infinity allows to extract the angular momentum $J$
\begin{equation}
\omega(r) = 
\frac{2J}{r^3}
+O(\frac{1}{r^5}) ,
\end{equation}
and the moment of inertia $I$, since $J=I \Omega$.
When calculating the moment of inertia for various EOSs, one obtains a large variation of its value for neutron stars with the same mass, as expected from the large variation of the radii, shown in Fig.~\ref{fig1} (see e.g. \cite{AltahaMotahar:2017ijw}).
This is illustrated in Fig.~\ref{fig2}, where the moment of inertia is shown versus the mass (Fig.~\ref{fig2}a) and the radius (Fig.~\ref{fig2}b) for several EOSs.

Of considerable interest is also the quadrupole moment $Q$ that is induced by rotation. 
However, to extract the quadrupole moment one has to go to second order in $\Omega$. 
As shown by Hartle and Thorne \cite{Hartle:1967he,Hartle:1968si,Pani:2014jra}
the appropriate parametrization of the metric is given by
\begin{eqnarray}
 ds^2 = &-& e^{2\Phi} \left[ 1 + 2 \epsilon^2 \left( h_0+h_2 P_2 \right) \right] dt^2   
 + e^{2 \Lambda} \left[ 1+ 2 e^{2 \Lambda} \epsilon^2 \left( m_0 + m_2 P_2 \right) / r  \right] dr^2 \nonumber \\
 &+& r^2 \left[ 1 +2 \epsilon^2 \left(v_2 -h_2 \right) P_2 \right] \left[ d\theta^2 + \sin^2 \theta \left( d\phi - \epsilon  \omega dt \right)^2 \right] ,
 \label{pert2}
\end{eqnarray}
where $P_2$ is the Legendre polynomial $P_2= \left( 3 \cos^2 \theta -1 \right) /2$, and $h_0$, $h_2$, $m_0$, $m_2$ and $v_2$ are radial functions. 
The density and pressure possess analogous second order terms.
After solving the resulting set of differential equations the quadrupole moment $Q$ can be read from the asymptotic behavior \cite{Hartle:1967he,Hartle:1968si,Pani:2014jra}
\begin{equation}
    h_2(r) \to \frac{Q}{r^3} .
\end{equation}

A further important property of neutron stars is their tidal deformability \cite{Hinderer:2007mb}.
In this case one considers a binary system, where the tidal forces of the companion compact object deform the neutron star \cite{Hinderer:2007mb,Yagi:2013awa,Pani:2014jra}.
The tidal Love number $\lambda$ is related to the tidal quadrupole moment and is obtained by placing the neutron star into an external quadrupolar tidal field.
The appropriate ansatz for the perturbations then consists of a subset of the previous ansatz for the rotational quadrupole moment with $\omega=h_0=m_0=0$ \cite{Pani:2014jra}.
The boundary conditions are of course different, since an external quadrupole field is present.
The asymptotic form of the function $h_2$,
\begin{equation}
   h_2 \to a_{-2} r^2 +a_{-1} r + \frac{a_3}{r^3} ,
\end{equation}
then provides the tidal Love number $\lambda$
\begin{equation}
    \lambda = \frac{a_3}{3 a_{-2}} .
\end{equation}
In a similar manner one can also obtain the higher multipole moments and the higher love numbers \cite{Yagi:2016bkt}.

\subsection{Quasi-Normal Modes}

Asteroseismology allows to extract important information on the stability and ringdown of neutron stars, when perturbed
(see e.g.~\cite{Andersson:1997rn,Kokkotas:1999bd}).
Neutron stars possess a rich spectrum of modes, associated with the nuclear matter and the gravitational field.
Since General Relativity features gravitational waves starting with quadrupole ($l=2$) radiation, QNMs arise when $l\ge 2$.
These possess a complex eigenvalue $\omega$ whose real part is the characteristic frequency $\omega_R$ of the mode, while the imaginary part $\omega_I$ represents its decay rate. 

The linear perturbations of the metric Ansatz and the fluid read \cite{Thorne:1967a}
\begin{eqnarray}
g_{\mu\nu} = g_{\mu\nu}^{(0)}(r) + \epsilon h_{\mu\nu}(t,r,\theta,\varphi)~, \\
\rho = \rho_0 (r) + \epsilon \delta\rho(t,r,\theta,\varphi)~,  \\
p = p_0(r) + \epsilon \delta p(t,r,\theta,\varphi)~, \\
u_{\mu} = u^{(0)}_{\mu} (r) + \epsilon \delta u_{\mu} (t,r,\theta,\varphi)
~,
\end{eqnarray}
where the superscript $(0)$ denotes the static and spherically symmetric background solutions.
The perturbations, in contrast, depend on all four coordinates. 

To proceed one then expands the perturbations in tensorial spherical harmonics characterized by multipole numbers $l$ and $m$ \cite{Thorne:1980ru}. 
The high symmetry of the background solutions then leads to a split of the perturbations into two separate classes: axial perturbations and polar perturbations. 
Axial perturbations transform as $(-1)^{l+1}$ under parity, and therefore do not couple to the fluid.
They are pure space-time modes.
Polar modes on the other hand are parity-even and transform as $(-1)^{l}$.
These include the perturbations of the pressure and energy density of the fluid.

Expansion and Fourier decomposition of the axial perturbations of the metric yields
\begin{equation}
h_{\mu\nu}^{(axial)} =  \sum\limits_{l,m} \int 
\left[
\begin{array}{c c c c}
	0 & 0 & - h_0	\frac{1}{\sin\theta}\frac{\partial}{\partial\phi}Y_{lm} 
	& h_0	\sin\theta\frac{\partial}{\partial\theta}Y_{lm} \\
  0 & 0 & - h_1	\frac{1}{\sin\theta}\frac{\partial}{\partial\phi}Y_{lm} 
  & h_1\sin\theta\frac{\partial}{\partial\theta}Y_{lm} \\
	-h_0\frac{1}{\sin\theta}\frac{\partial}{\partial\phi}Y_{lm} & 
	- h_1\frac{1}{\sin\theta}\frac{\partial}{\partial\phi}Y_{lm}  & 0 & 0 \\
	h_0	\sin\theta\frac{\partial}{\partial\theta}Y_{lm} & 
	h_1	\sin\theta\frac{\partial}{\partial\theta}Y_{lm} & 0 & 0
\end{array}
\right]
e^{-i\omega t} d\omega ,
\label{pert1}
\end{equation}
whereas for polar perturbations one finds
\begin{equation}
h_{\mu\nu}^{(\text{polar})} = \sum\limits_{l,m}\,\int    
\left[
\begin{array}{c c c c}
r^l e^{2\nu} H_0 Y_{lm} & -i \omega r^{l+1} H_1 Y_{lm} & 
0 & 0 \\
-i \omega r^{l+1} H_1 Y_{lm} &  r^l e^{2\lambda} H_2 Y_{lm} & 
0 & 0 \\
0 & 0
& r^{l+2} K  Y_{lm} & 0
\\
0 & 0  & 
0 & r^{l+2}  \sin^2\theta K  Y_{lm} \\
\end{array}
\right]
e^{-i\omega t} d\omega
~,
\end{equation}
(using $(t,r,\theta,\varphi)$ order for the matrix).
The corresponding decomposition of the density and the pressure of the fluid inside the star is
\begin{equation}
\delta \rho = \sum\limits_{l,m}\,
\int   r^l E_{1 } Y_{lm} e^{-i\omega t }d\omega~, \quad
\delta p = \sum\limits_{l,m}\,
\int   r^l \Pi_{1}  Y_{lm} e^{-i\omega t} d\omega~, 
\end{equation}
and the perturbation of the velocity is
\begin{eqnarray}
\delta u_{\mu} = 
\sum\limits_{l,m}\,\int    
\left[
\begin{array}{c}
\frac{1}{2} r^l e^{\nu} H_0 Y_{lm}  \\
r^l i\omega e^{-\nu} 
\left(e^{\lambda}W/r -r H_1 \right) Y_{lm}  \\
-i\omega r^l e^{-\nu} V \partial_{\theta} Y_{lm}
\\
-i\omega r^l e^{-\nu} V \partial_{\phi} Y_{lm}  \\
\end{array}
\right]
e^{-i\omega t} d\omega
~,
\end{eqnarray}
while outside the star there is no fluid, of course.
All perturbation functions depend only on the radial coordinate $r$, the multipole numbers $l$, $m$, and the complex eigenvalue $\omega$.
The resulting systems of ordinary differential equations must then be simplified by specific choices of gauge 
and solved subject to an appropriate set of boundary conditions.
These boundary conditions require regularity at the center of the star and a purely outgoing wave behavior at infinity.
Moreover, they require continuity of the metric perturbation functions and their derivatives at the border of the star, where the pressure and the energy density vanish.
Together all these requirements then select a discrete set of values for the complex eigenvalue $\omega$ for a given $l$, representing the fundamental frequency and its overtones (see e.g. \cite{Lindblom:1983ps,Detweiler:1985zz} for further details).

\begin{figure}[ht]
\begin{center}
\mbox{\hspace{-0.3cm}
\includegraphics[width=.51\textwidth, angle =0]{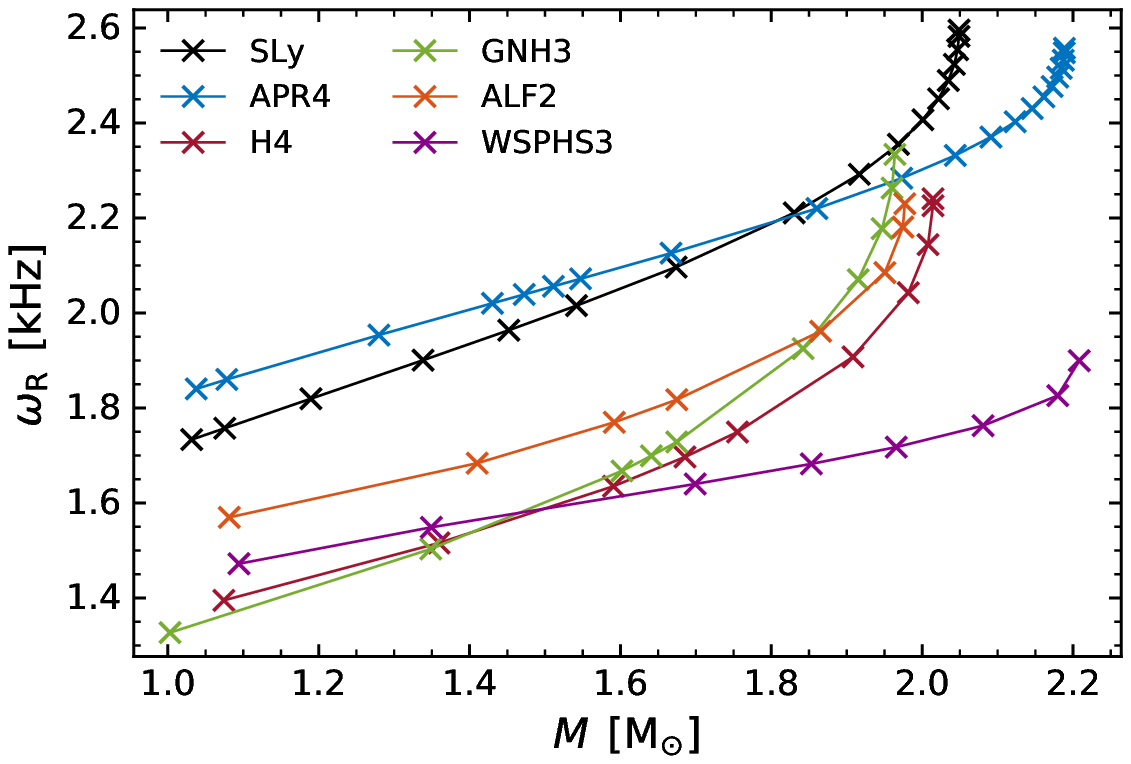}
\includegraphics[width=.51\textwidth, angle =0]{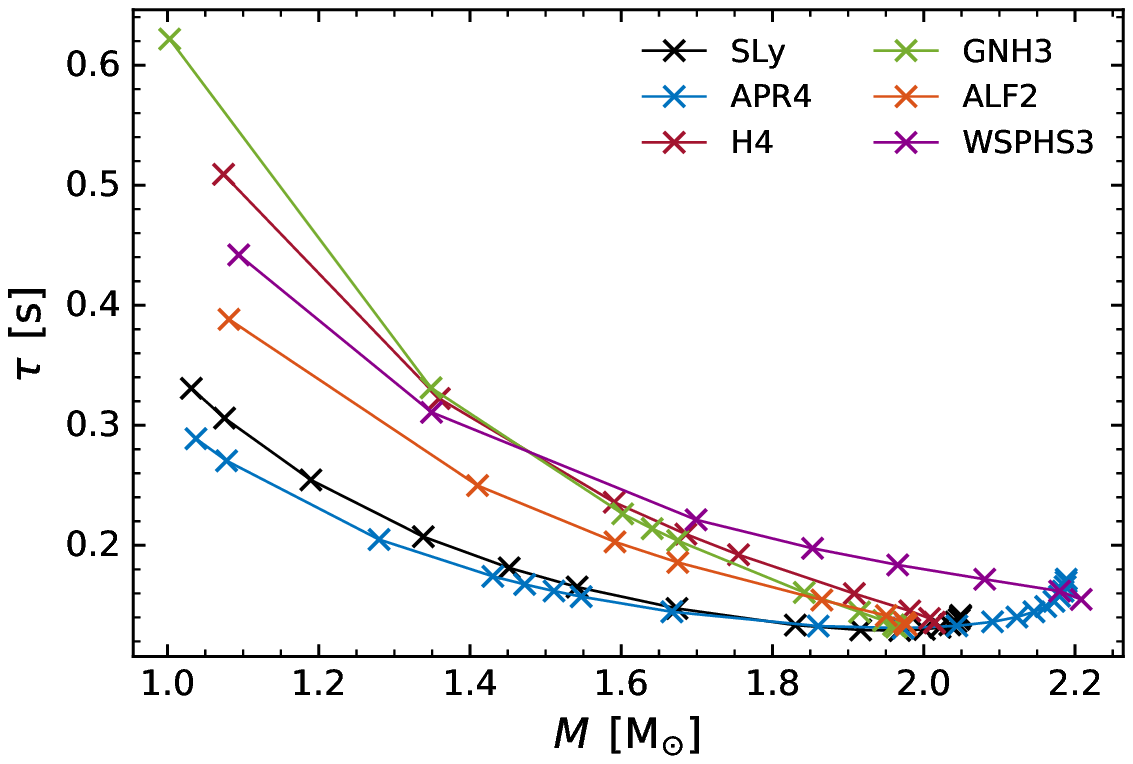}
}
\end{center}
\vspace{-0.5cm}
\caption{{\it{Fundamental $f$ mode ($l=2$) of neutron stars in GR: (a) frequency $\omega_R$ (in kHz) vs mass $M$ (in solar masses $M_\odot$); (b) decay time $\tau=1/\omega_I$ (in s) vs mass $M$ (in solar masses $M_\odot$) for several EOSs. 
}}}
\label{fig3}
\end{figure}

The fundamental $f$ mode (l=2) in GR is illustrated in Fig.~\ref{fig3}, where the frequency $\omega_R$ (Fig.~\ref{fig3}a) and the decay time $\tau=1/\omega_I$ (Fig.~\ref{fig3}a) are shown versus the mass of the neutron stars for several EOSs. The figure reveals clearly the significant dependence of the modes on the EOS. 

\section{Universal Relations}

As discussed above, dimensionful neutron star properties depend significantly on the employed EOS.
If, however, properly scaled dimensionless quantities are considered instead, an important set of \textit{universal relations} arises in GR, which exhibit only little EOS dependence.

\subsection{$I$-Love-$Q$ Relations}

In geometric units the so-called compactness $C$ is a simple dimensionless quantity.
It represents the ratio of the mass $M$ and the radius $R$ of a neutron star, $C=M/R$.
The compactness of neutron stars ranges typically in the interval $0.1<C<0.3$, while a Schwarzschild black hole has a compactness of $C=0.5$, since its horizon radius is given by $R=2M$.
Clearly, compactness is a relevant physical property, and being dimensionless, it is a suitable candidate to feature in \textit{universal relations}.

\begin{figure}[ht]
\begin{center}
\mbox{\hspace{-0.3cm}
\includegraphics[width=.51\textwidth, angle =0]{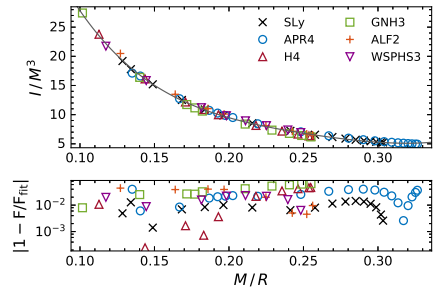}
}
\end{center}
\vspace{-0.5cm}
\caption{{\it{Universal $\bar I$-$C$ relation for several EOSs. 
}}}
\label{fig4}
\end{figure}

A first \textit{universal relation} can thus be envisaged that exhibits a suitably scaled moment of inertia $\bar I$ versus the compactness $C$.
Since $J=\Omega I$, a dimensionless moment of inertia is obtained in geometric units in terms of $\bar I=I/M^3$. (Recall, that $J/M^2$ is dimensionless.)
This $\bar I$-$C$ relation is demonstrated in Fig.~\ref{fig4} for several EOSs.
While dependence on the EOS has been reduced considerably for these dimensionless quantities as compared to the dimensionful quantities $I$, $M$ and $R$ shown in Fig.~\ref{fig2}, this relation is less impressive than the $I$-Love-$Q$ relations discussed in the following.

Besides the dimensionless moment of inertia $\bar I$ the dimensionless quadrupole moment $\bar Q=QM/J^2$ and the dimensionless Love number $\bar \lambda=\lambda/M^5$ feature prominently in the $I$-Love-$Q$ relations. 
These relations do not involve the compactness, but consider only the dimensionless quantities $\bar I$, $\bar \lambda$ and $\bar Q$.
Obtained by Yagi and Yunes \cite{Yagi:2013bca,Yagi:2016bkt}, the truly remarkable $I$-Love, $I$-$Q$, and $Q$-Love relations can be expressed as simple curves of the type 
\begin{equation}
\ln y_i 
= a_i + b_i \ln x_i + c_i (\ln x_i)^2 + d_i (\ln x_i)^3+ e_i (\ln x_i)^4 ,
\end{equation}
where $y_i$ represent the first and $x_i$ the second dimensionless quantity, as seen in Table I, where the coefficients $a_i$ to $e_i$ yield an excellent fit to the data of a very large number of EOSs with very different properties of the matter of the star \cite{Yagi:2013bca,Yagi:2016bkt}. In fact, the deviations of the data from the best fit shown are below 1\%.

\begin{center}
\begin{tabular}{cccccccc}
\hline
\hline
\noalign{\smallskip}
$y_i$ & $x_i$ &&  \multicolumn{1}{c}{$a_i$} &  \multicolumn{1}{c}{$b_i$}
&  \multicolumn{1}{c}{$c_i$} &  \multicolumn{1}{c}{$d_i$} &  \multicolumn{1}{c}{$e_i$}  \\
\hline
\noalign{\smallskip}
$\bar{I}$ & $\bar{\lambda}$ && 1.496 & 0.05951  & 0.02238 & $-6.953\times 10^{-4}$ & $8.345\times 10^{-6}$\\
$\bar{I}$ & $\bar Q$ && 1.393  & 0.5471 & 0.03028  & 0.01926 & $4.434 \times 10^{-4}$\\
$\bar Q$ & $\bar{\lambda}$ && 0.1940  & 0.09163 & 0.04812  & $-4.283 \times 10^{-3}$ & $1.245\times 10^{-4}$\\
\noalign{\smallskip}
\hline
\hline
\end{tabular}
\end{center}

Analogous relations can be considered for the higher multipole moments and higher Love numbers.
In the usual nomenclature the mass corresponds to the lowest mass moment $M=M_0$ and the angular momentum to the lowest current moment $J=S_1$.
Higher mass moments possess even index, $M_{2l}$, and higher current moments odd index, $S_{2l+1}$  \cite{Geroch:1970cd,Hansen:1974zz,Thorne:1980ru,Hoenselaers:1992bm,Sotiriou:2004ud}.
The quadrupole moment then corresponds to $Q=M_2$.
Higher tidal mass and current moments are referred to as
$\lambda_n$ \cite{Damour:2009vw}.
Examples of \textit{universal relations} for higher moments are exhibited in Table II \cite{Stein:2013ofa,Yagi:2016bkt}. 
These represent the $\bar S_3$-$\bar Q$ and $\bar M_4$-$\bar Q$ relations, that possess larger deviations (4\% and 10\%) than the $I$-Love-$Q$ relations.

\begin{center}
\begin{tabular}{cccccccc}
\hline
\hline
\noalign{\smallskip}
$y_i$ & $x_i$ &&  \multicolumn{1}{c}{$a_i$} &  \multicolumn{1}{c}{$b_i$}
&  \multicolumn{1}{c}{$c_i$} &  \multicolumn{1}{c}{$d_i$} &  \multicolumn{1}{c}{$e_i$}  \\
\hline
\noalign{\smallskip}
$\bar S_3$ & $\bar Q$ && $3.131 \times 10^{-3}$  & 2.071 & $-0.7152$  & 0.2458 & $-0.03309$\\
$\bar M_4$ & $\bar Q$ && $-0.02287$  & 3.849 & $-1.540$  & 0.5863 & $-8.337\times 10^{-2}$\\
\noalign{\smallskip}
\hline
\hline
\end{tabular}
\end{center}

In this connection the expression \textit{three hair relations} was coined \cite{Stein:2013ofa,Yagi:2016bkt}. 
This is a generalization of the \textit{no-hair} (\textit{two hair relation}), highlighting that Kerr black holes are fully determined by only two quantities (hairs), their mass and their angular momentum.
In the \textit{three hair relations} of neutron stars the additional quantity besides the mass and the angular momentum is the quadrupole moment \cite{Stein:2013ofa,Yagi:2016bkt}.
In contrast to the \textit{two hair relation} of black holes, the \textit{three hair relations} of neutron stars are only approximate relations.
Their validity for neutron stars has been associated with an approximate symmetry that emerges at high compactness: the self-similarity of isodensity surfaces \cite{Yagi:2014qua,Yagi:2016bkt}.

\subsection{Quasi-Normal Modes}

\textit{Universal relations} arise also in the study of quasi-normal modes.
As illustrated above, quasi-normal modes feature a considerable dependence on the EOS.
However, Anderson and Kokkotas pointed out rather early that \textit{universal relations} may reduce this EOS dependence significantly \cite{Andersson:1996pn,Andersson:1997rn}, as confirmed in numerous further studies, e.g., \cite{Kokkotas:1999mn,Benhar:1998au,Benhar:2004xg,Tsui:2004qd,Lau:2009bu,Blazquez-Salcedo:2012hdg,Blazquez-Salcedo:2013jka,Chirenti:2015dda}.

\begin{figure}[ht]
\begin{center}
\mbox{\hspace{-0.3cm}
\includegraphics[width=.51\textwidth, angle =0]{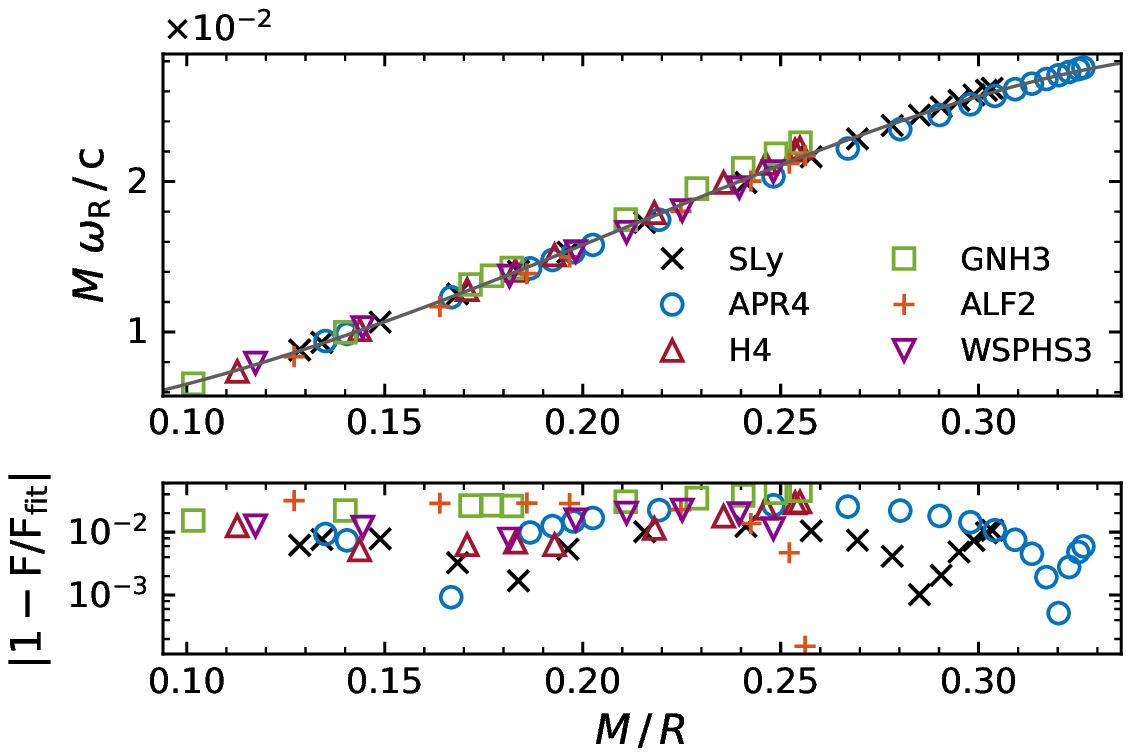}
\includegraphics[width=.51\textwidth, angle =0]{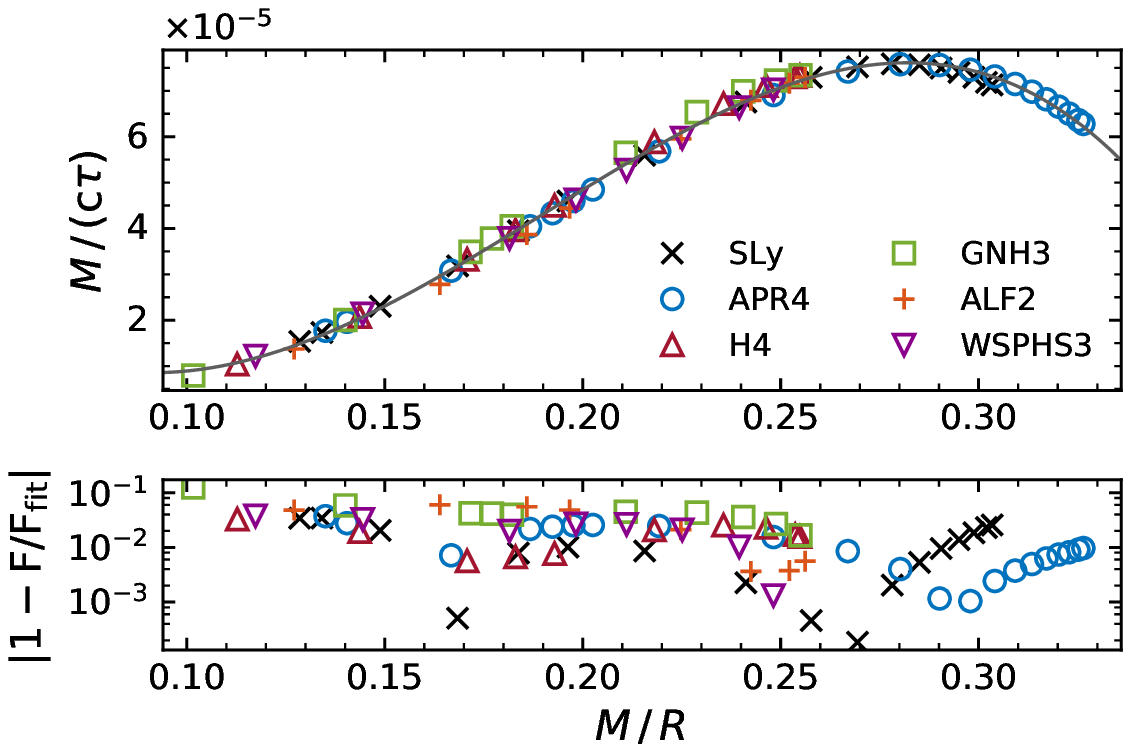}
}
\mbox{\hspace{-0.3cm}
\includegraphics[width=.51\textwidth, angle =0]{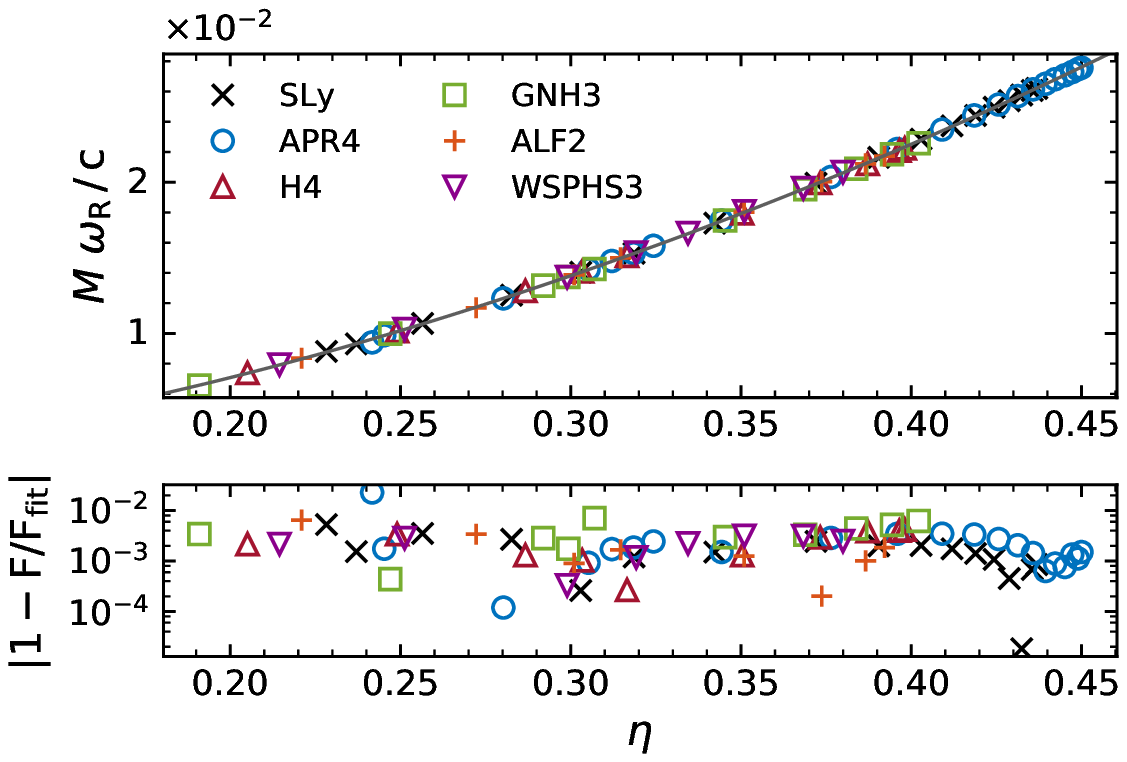}
\includegraphics[width=.51\textwidth, angle =0]{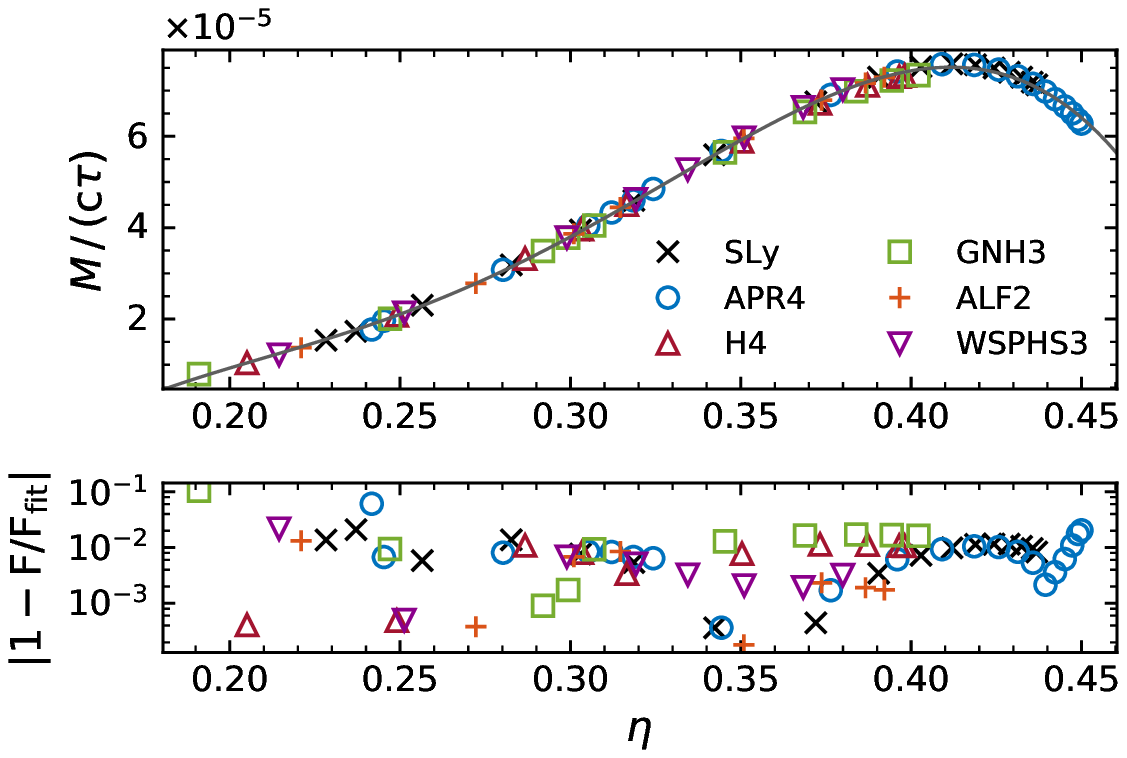}
}
\end{center}
\vspace{-0.5cm}
\caption{{\it{Universal relation for the fundamental $f$ mode ($l=2$) of neutron stars in GR: (a) scaled frequency $M\omega_R$/c vs compactness $C=M/R$; (b) scaled decay rate $M\omega_I/c$ vs compactness $C=M/R$ for several EOSs;
(c) and (d) analogous, but vs the effective compactness $\eta=\sqrt{M^{3}/I}$. 
}}}
\label{fig5}
\end{figure}

A set of \textit{universal relations} for the fundamental $f$ mode of neutron stars is illustrated in Fig.~\ref{fig5}, where Fig.~\ref{fig5}a and Fig.~\ref{fig5}b exhibit the dimensionless scaled frequency $M\omega_R/c$ and the dimensionless scaled decay rate $M\omega_I/c$, respectively, versus the compactness $C=M/R$ for several EOSs.
Analogous relations are shown in Fig.~\ref{fig5}c and Fig.~\ref{fig5}d, where instead of the compactness $C$ the so-called effective compactness $\eta=\sqrt{M^{3}/I}={\bar I}^{-1/2}$ was used, that is based on the dimensionless moment of inertia $\bar I$.
It was introduced in \cite{Lau:2009bu}, where also the following best fit was provided for the $f$ mode
\begin{equation}
M \omega_R = -0.0047 + 0.133 \eta + 0.575 \eta^2
, \ \ \  M \omega_I  = 0.00694 \eta^4 - 0.0256 \eta^6 .
\nonumber
\end{equation}
The parametrization in terms of the effective compactness reduces the errors (as compared to the compactness) and is therefore preferable.
Various further \textit{universal relations} for the quasi-normal modes have been found, among them for instance a relation between the scaled frequency of a mode and the scaled damping rate \cite{Blazquez-Salcedo:2012hdg,Blazquez-Salcedo:2013jka}.

\textit{Universal relations} can be of use in many different circumstances \cite{Yagi:2016bkt}.
First of all, they can be employed to extract further information on neutron star properties not yet known from explicit measurements.
In case of the lowest moments, for instance, the $I$-Love-$Q$ relations would allow to obtain any two of the three quantities, once the third one would be measured \cite{Yagi:2016bkt}, while the seven lowest moments could be obtained from measurements of the mass, rotation period and moment of inertia with the help of the \textit{three-hair-relations} \cite{Stein:2013ofa}.
On the other hand, in the case of the quasi-normal mode measurement of an axial or polar mode would allow the determination of the mass $M$ and the moment of inertia $I$ of a star by invoking the $M\omega_R$-$\eta$ and $M\omega_I$-$\eta$ relations \cite{Lau:2009bu}. 
Moreover, the radius $R$ might be extracted and conclusions with respect to the EOS might be possible.
Currently \textit{universal relations} are already employed to reduce degeneracies in the analysis of gravitational waves \cite{Yagi:2016bkt}.
Last but not least, as discussed next \textit{universal relations} also provide a means to test alternative gravity theories.

\section{Neutron Stars in Alternative Gravity Theories}

Studies of alternative gravity theories are motivated largely by the quest for a theory of quantum gravity and by cosmological issues like dark matter and dark energy.
Such theories typically involve new degrees of freedom, with the simplest being a real scalar field.
If indeed such additional degrees of freedom would be present, their consequences might not only resolve the issues intended, but they might also have observable consequences that could be tested by observations in the solar system or observations of black holes and neutron stars and gravitational waves emitted by these compact objects \cite{Will:2005va,Capozziello:2010zz,Berti:2015itd,CANTATA:2021ktz}.
In the following neutron stars will be discussed for two widely employed types of alternative gravity theories. 

\subsection{Scalar-Tensor-Theories}

Scalar-tensor theories introduce in addition to the gravitational tensor field a gravitational scalar field (see e.g.~\cite{Brans:1961sx,Damour:1992we,Damour:1996ke,Fujii:2003pa}).
A generic action for such scalar-tensor theories is given by
\begin{equation}
S = \frac{1}{16\pi G} \int d^4x \sqrt{-{\tilde
g}}\left[{F(\Phi)\tilde {\cal R}} - Z(\Phi){\tilde
g}^{\mu\nu}\partial_{\mu}\Phi
\partial_{\nu}\Phi   -2 U(\Phi) \right] +
S_{m}\left[\Psi_{m};{\tilde g}_{\mu\nu}\right] ,
\label{action_Jordan}
\end{equation}
where the tilde indicates that the respective quantities are in the so-called Jordan frame, $\Phi$ is the gravitational scalar field, and $S_m$ denotes any additional matter fields $\Psi_{m}$.
In the Jordan frame the gravitational scalar field does not couple directly to the matter fields and the weak equivalence principle is retained.
The functions $F(\Phi)$ and $Z(\Phi)$ cannot be chosen arbitrary, but need to meet some physical restrictions \cite{Esposito-Farese:2000pbo}.

While neutron stars can be studied directly in the Jordan frame, it is typically more convenient to transform to the so-called Einstein frame, which can be achieved by means of a conformal transformation of the metric $g_{\mu\nu} = F(\Phi){\tilde g}_{\mu\nu}$, 
and an associated transformation of the gravitational scalar field denoted by $\varphi$ now \cite{Damour:1992we,Damour:1996ke,Fujii:2003pa}.
\begin{equation}
\left( \frac{d\varphi}{d\Phi} \right)^2 = 
\frac{3}{4}\left( \frac{d\ln(F(\Phi))}{d\Phi} \right)^2 
+ \frac{Z(\Phi)}{2 F(\Phi)} .
\label{scalar_trans}
\end{equation}
In the Einstein frame, the action then reads
\begin{equation}
S= \frac{1}{16\pi G}\int d^4x \sqrt{-g} \left[{\cal R} -
2g^{\mu\nu}\partial_{\mu}\varphi \partial_{\nu}\varphi -
4V(\varphi)\right]+ S_{m}[\Psi_{m}; \mathrm{A}^{2}(\varphi)g_{\mu\nu}] ,
\label{action_Einstein}
 \end{equation}
where the Einstein frame quantities are denoted without tilde, and the following relations hold 
\begin{equation}
{A}(\varphi) = F^{-1/2}(\Phi) \, , \,\, 2V(\varphi) = U(\Phi)F^{-2}(\Phi) .
\end{equation}
In the simplest case the scalar potential is chosen to vanish, $U(\Phi)=0=V(\varphi)$.
The gravitational scalar field is then massless and has no self-interactions.

Variation of the action (\ref{action_Einstein}) leads to the Einstein equations
\begin{equation} 
{\cal R}_{\mu\nu} - \frac{1}{2}g_{\mu\nu}{\cal R} =
  2\partial_{\mu}\varphi \partial_{\nu}\varphi   -
g_{\mu\nu}g^{\alpha\beta}\partial_{\alpha}\varphi
\partial_{\beta}\varphi
+ 8\pi T_{\mu\nu} 
\end{equation}
and gravitational scalar field equation
\begin{equation}
 \nabla^{\mu}\nabla_{\mu}\varphi = - 4\pi k(\varphi)T,
\end{equation}  
where $T = T^{\mu}_{\mu}$, and $k(\varphi)= \frac{d\ln({A}(\varphi))} {d\varphi}$. 
The function $A(\varphi)$ determines the coupling between the scalar field and the matter.
The stress-energy tensor ${\tilde T}_{\mu\nu}$ is provided in the physical Jordan frame and then transformed into the Einstein frame
\begin{equation}
T_{\mu\nu}= {A}^2 {\tilde T}_{\mu\nu} ,
\end{equation}
where the Bianchi identities yield
\begin{equation}
\nabla_{\mu}T^{\mu}{}_{\nu} = k(\varphi)T\partial_{\nu}\varphi . \end{equation}

The freedom in the choice of coupling function $A(\varphi)$ leads to different types of scalar-tensor theories, and thus different consequences for neutron stars in these theories.
At the same time it leads to a variety of physical effects, that can be compared to observations and thus result in more or less stringent constraints from observations.
Brans-Dicke theory, for instance, is obtained for the simple parametrization $A=e^{\kappa \varphi}$, i.e., $k(\varphi)=\kappa$ with constant $\kappa$, addressed further below \cite{Brans:1961sx}.
Here another coupling function is considered,
\begin{equation}
{A}(\varphi)=e^{\frac{1}{2}\beta\varphi^2} \ , \ \ \ k(\varphi)=\beta \varphi ,
\label{A1}
\end{equation}
that leads to the interesting phenomenon of spontaneous scalarization of neutron stars, discovered by Damour and Esposito-Far\`ese \cite{Damour:1993hw}.
 
Spontaneous scalarization in neutron stars is matter induced.
It can arise in theories with coupling functions, that possess a quadratic dependence on the gravitational scalar field such that it satisfies a Klein-Gordon type equation with an effective mass, i.e.,
\begin{equation}
 \nabla^{\mu}\nabla_{\mu}\varphi = m^2_{\rm eff} \varphi .
\end{equation} 
In that case the GR neutron star solutions remain solutions of the scalar-tensor theory, since for vanishing scalar field the equations reduce to the GR equations.
However, in addition to the GR solutions new solutions with a gravitational scalar field may arise, when the neutron matter represents a sufficiently strong source to induce a tachyonic instability, $m_{\rm eff}^2 = - 4\pi G \beta T < 0$. 
While typical neutron stars possess $T=3p-\rho<0$, so $\beta<0$ must be chosen for spontaneous scalarization to occur, both $T$ and $\beta$ could also be positive, but in this case the neutron stars would need a pressure dominated core \cite{Mendes:2014ufa,Mendes:2016fby}.

When evaluating a family of GR neutron stars by increasing the central pressure, at some point a neutron star with a zero mode  arises.
Beyond this point scalarization sets in, and a branch of scalarized neutron stars is present in addition to the GR neutron stars.
In fact, GR neutron stars then possess an unstable mode, whereas the scalarized neutron stars become the physically preferred stable configurations (see, e.g., \cite{Damour:1993hw,Mendes:2018qwo}).
The scalar field at the center of the star and the scalar charge are largely independent of the EOS, and thus basically universal, only depending on the gravitational potential at the center of the neutron star \cite{AltahaMotahar:2017ijw}.

Pulsar observations have by now virtually excluded the possibility of spontaneous scalarization of neutron stars for the simplest case of a massless scalar field \cite{Zhao:2022vig}.
These conclusions are based on the expected effects of dipolar and thus scalar radiation on the orbits of the compact objects. 
However, the inclusion of a genuine mass term with a sufficiently large mass for the scalar field allows to circumvent these observational constraints, since the dipolar radiation becomes rather negligible when the orbital separation of a binary star system is much larger than the scalar field Compton wavelength \cite{Ramazanoglu:2016kul}.
Evaluation of the properties and quasi-normal modes of scalarized neutron stars with a massive gravitational scalar field also leads to \textit{universal relations}
(see e.g., \cite{Yazadjiev:2016pcb,Doneva:2016xmf,AltahaMotahar:2019ekm}).
Depending on the strength of the scalarization, they may differ significantly from those of neutron stars in GR.

\subsection{$f({\cal R})$ Theories}

In $f({\cal R})$ theories the gravitational action is no longer given by the curvature scalar ${\cal R}$, but by some function of the curvature scalar, $f({\cal R})$ \cite{Sotiriou:2008rp,DeFelice:2010aj,Capozziello:2011et}.
A particular well-motivated such theory is based on
\begin{equation}
    f({\cal R})={\cal R}+a{\cal R}^2 \ .
    \label{R2}
\end{equation}
$f({\cal R})$ theories can also be transformed to the Einstein frame, since from a mathematical point of view they are equivalent to scalar-tensor theories.
As shown in \cite{Yazadjiev:2014cza,Staykov:2014mwa} such a transformation then leads to a scalar field with a Brans-Dicke type coupling function and a scalar field potential
\begin{equation}
    A(\varphi)= e^{-\frac{1}{\sqrt{3}}\varphi}  \ , \ \ \ V(\varphi)=\frac{3m_{\varphi}^2}{2} \big(1- e^{-\frac{2\varphi}{\sqrt{3}}}\big)^2 \ .
\end{equation}
The scalar field mass $m_\varphi$ is identified in the transformation from the coefficient of the $\varphi^2$ term of the potential $V(\varphi)$, that arises in the transformation, and is thus a function of the coupling constant $a$ of the $f(R)$ theory considered, $m_\varphi=1/\sqrt{6a}$. 
The parameter $a$ therefore determines the mass of the scalar field, and can be chosen well within the current observational window \cite{Ramazanoglu:2016kul}.

Besides leading to distinct $I$-Love-$Q$ relations (see e.g., \cite{Yazadjiev:2015zia,Doneva:2015hsa}), this $f({\cal R})$ theory has a distinct spectrum of quasi-normal modes \cite{Blazquez-Salcedo:2018qyy,Blazquez-Salcedo:2020ibb,Blazquez-Salcedo:2021exm,Blazquez-Salcedo:2022}.
In particular, in contrast to GR, monopole ($l=0$) and dipole ($l=1$) radiation arises due to the additional degree of freedom.
In GR neutron stars possess only $l=0$ normal modes.
But in such an $f({\cal R})$ theory, these modes become propagating modes.
Interestingly, these modes feature a very small decay rate $\omega_I$, which means that they are ultra long-lived \cite{Blazquez-Salcedo:2020ibb}.
Moreover, the scale of the frequency $\omega_R$ is determined by the neutron star size for small Compton wavelength $L_\varphi=1/m_\varphi$ of the scalar field, while for large $L_\varphi$ the frequency follows $L_\varphi$.

\begin{figure}[ht]
\begin{center}
\mbox{\hspace{-0.3cm}
\includegraphics[width=.51\textwidth, angle =0]{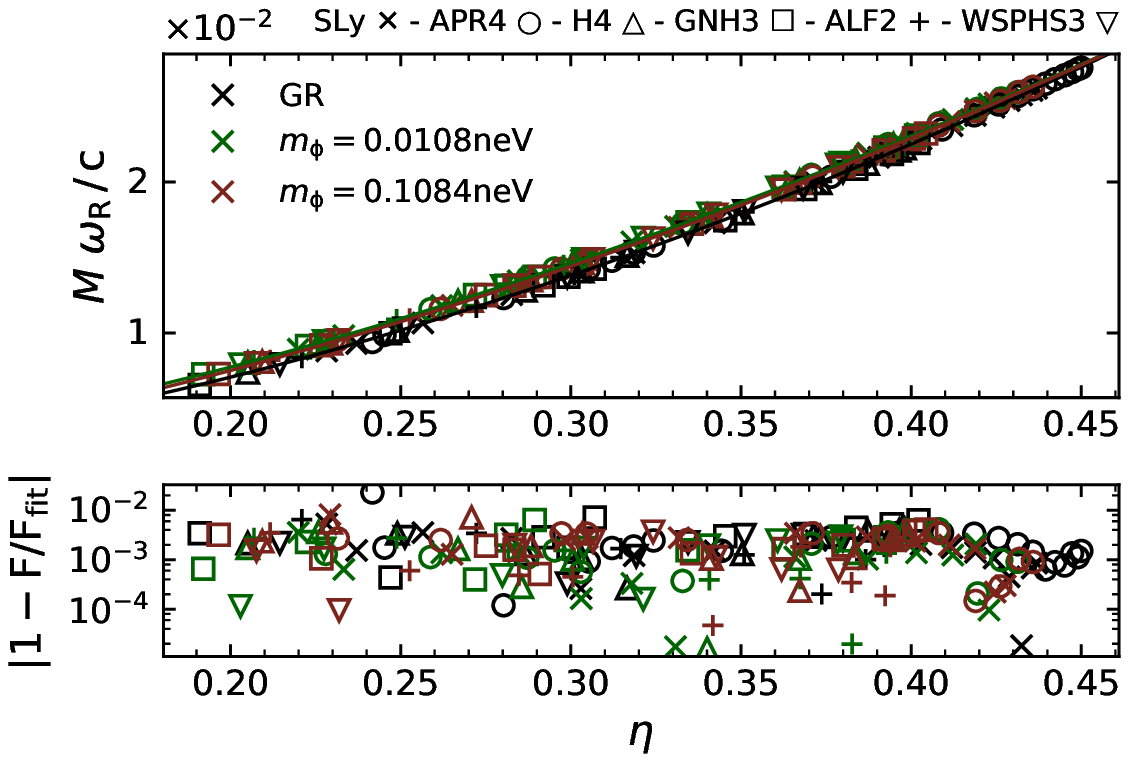}
\includegraphics[width=.51\textwidth, angle =0]{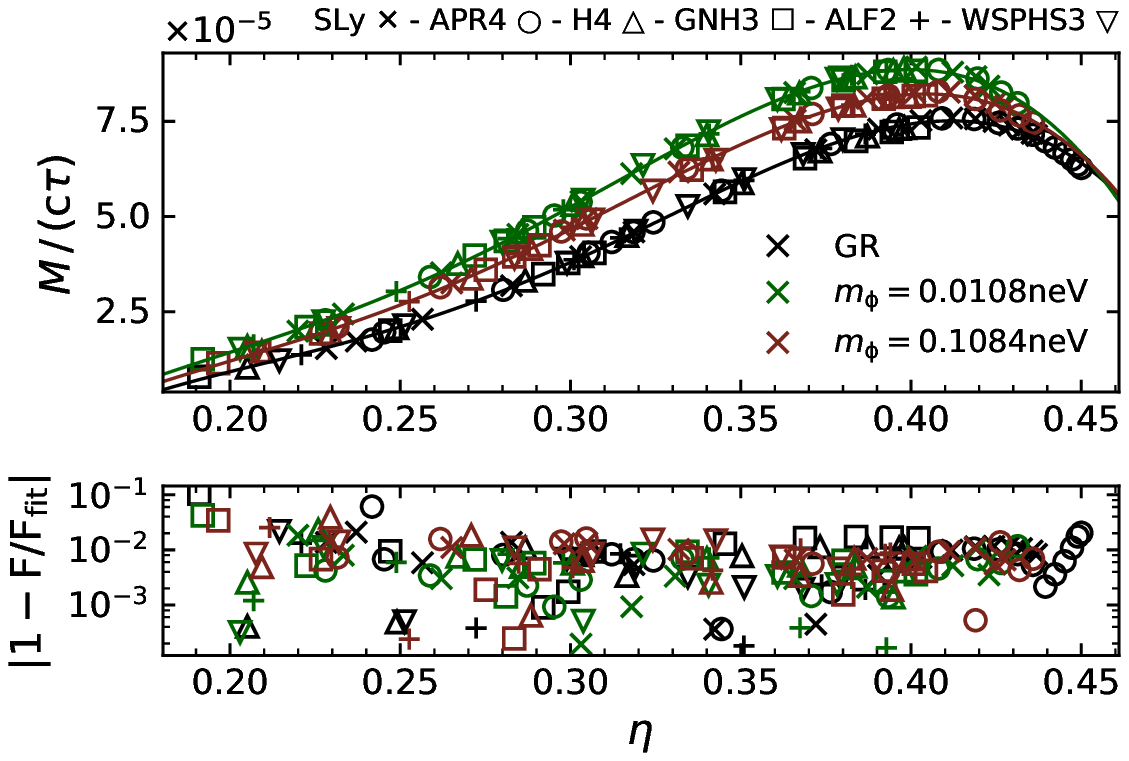}
}
\end{center}
\vspace{-0.5cm}
\caption{{\it{Universal relation for the fundamental $f$ mode ($l=2$) of neutron stars in $f({\cal R})={\cal R}+a{\cal R}^2$ theory with $m_\varphi=0.0108$ neV and 0.1084 neV: (a) scaled frequency $M\omega_R$/c vs effective compactness $\eta=\sqrt{M^{3}/I}$  (b) scaled decay rate $M\omega_I$/c vs effective compactness $\eta=\sqrt{M^{3}/I}$ for several EOSs. For comparison also the GR relations are shown.
}}}
\label{fig6}
\end{figure}

The \textit{universal relations} for quasi-normal modes exhibit distinct features, as well, and therefore might be exploited to put further bounds on such a theory \cite{Blazquez-Salcedo:2018qyy,Blazquez-Salcedo:2022}.
Fig.~\ref{fig6} illustrates a set of \textit{universal relations} for the fundamental f mode ($l=2$) for two values of the scalar mass $m_\varphi$ for this $f({\cal R})$ theory, analogous to fig.~\ref{fig5} for GR.
Due to the presence of the new degree of freedom, however, this f mode is not the only polar quadrupole ($l=2$) mode.
There is an additional scalar $l=2$ mode present, and there are also the scalar dipole and monopole modes, all of them exhibiting \textit{universal relations} \cite{Blazquez-Salcedo:2022}.

\section{Conclusion}

Since neutron stars are highly compact objects, they represent ideal astrophysical objects to learn about gravity.
While the current lack of knowledge of the physical EOS of nuclear matter under these extreme conditions leads to larger ranges of possible values of their physical properties and their emitted radiation, the presence of \textit{universal relations}, which are rather independent of the EOS, reduces these uncertainties to a large extent.
Moreover, \textit{universal relations} may be rather different for GR and for alternative theories of gravity, thus allowing to put bounds on such theories once the corresponding measurements will have been achieved with sufficient accuracy.
\\

\textbf{Acknowledgments}. {I would like to thank the organizers for the invitation to the interesting meeting \textsl{Signatures and experimental searches for modified and quantum gravity}.
I would also like to thank my collaborators and here, in particular, Jose Luis Bl\'azquez-Salcedo and Vincent Preut, for providing the above figures. Furthermore, I would like to gratefully acknowledge support by the DFG Research Training Group 1620 \textit{Models of Gravity} and the COST Actions CA15117 and CA16104.}

\end{document}